\documentclass[twocolumn,aps,showpacs,amsmath,amssymb]{revtex4}

\usepackage{graphicx}
\usepackage{dcolumn}
\usepackage{bm}
\usepackage{epsfig}
\usepackage{citesort}
\usepackage{times}
\begin{document}

\title{Density of states of colloidal glasses}
\author{Antina Ghosh$^1$}
\author{Vijayakumar K. Chikkadi$^1$}
\author{Peter Schall$^1$}
\author{Jorge Kurchan $^2$}
\author{Daniel Bonn$^{1,3}$}
\affiliation{$^{1}$Van der Waals Zeeman Institute,
University of Amsterdam, Valckenierstraat 67, 1018 XE Amsterdam The Netherlands.\\
$^2$PMMH, ESPCI, 10 rue Vauquelin, CNRS UMR 7636 , Paris, France 75005,\\
$^3$LPS de l'ENS, CNRS UMR 8550, 24 Rue Lhomond Paris, France 75005.}

\date{\today}

\begin{abstract}\textbf{
Glasses are structurally liquid-like, but mechanically solid-like.
Most attempts to understand glasses start from liquid state
theory. Here we take the opposite point of view, and use concepts
from solid state physics.  We determine the  vibrational
modes of a colloidal glass experimentally,
and find  soft low-frequency modes that are
very different in nature from the usual acoustic vibrations of
ordinary solids. These modes extend over surprisingly large length scales. }
\end{abstract}

\pacs{47.27.Cn, 47.20.Ft, 47.50.+d}
\maketitle
The glass transition is perhaps the greatest unsolved problem in
condensed matter physics \cite{Hansen}: the main question is how to reconcile
the liquid-like structure with solid-like mechanical properties.
In solids, structure and mechanics are related directly through
the vibrational density of states (DOS) of the material
\cite{kittel}. 
It then seems important to obtain directly 
the density of states of a glass that is solid mechanically, but
has no crystalline ordering. 
We do so for (colloidal) hard spheres that are known to undergo a
glass transition upon increasing volume fraction.

Recent theory shows that randomly packed hard spheres form a system with
no redundant mechanical constraints. Because of this, a small
perturbation, like for example breaking a particle contact, may
induce rearrangements at all scales~\cite{alexander,combe,Cooper}:
the system is `critical' in this sense~\cite{silbert,WW}. It has
been argued that the marginality of the system has dramatic
consequences for the vibrational spectrum ~\cite{Wyart}: in stark
contrast to well-known phonon modes in solids, which have been
observed in previous studies \cite{Keim,Noel,Water,Chaikin},  a
broad band of floppy modes emerges in the vibrational density of
states, with a gap at low frequencies that disappears  as the
pressure becomes infinite.

In this Letter we measure, for the fist time, vibrational eigenstates
of a colloidal glass/supercooled liquid  system.  We consider
colloidal hard spheres that are subject to thermal agitation,
allowing us to follow the  `vibrational' motion of particles that
are trapped in cages constituted of neighboring particles. This is
done at different random dense packing configurations around the
glass transition. We obtain information on the  DOS from a
normal-mode analysis of particle displacements measured using
confocal microscopy. We find that the vibrational spectrum has
many soft low-frequency modes \cite{Wyart}, more abundant and very
different in nature from the usual acoustic vibrations of ordinary
solids. This results in an anomalous low frequency peak in the
density of states which approaches zero frequency as one goes
deeper into the glass phase. The observed soft modes are
collective 'swirling' particle motions that extend over
surprisingly long length scales.

\begin{figure}[b]
\begin{center}
\includegraphics[width=5cm,height=5cm]{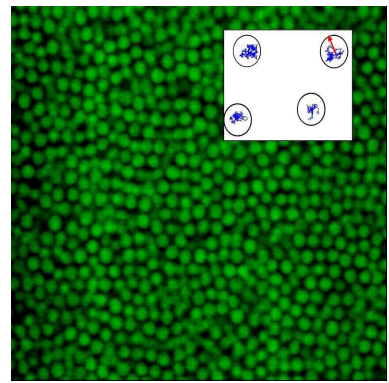}
\caption{A two dimensional image of the colloidal suspension at a
volume fraction $0.60$ acquired by confocal microscopy. 
The inset shows a few typical particle trajectories,
from which it is clear that we obtain the DOS before large-scale
rearrangements happen.} \label{Fig_2Dimage} \label{Fig_2Dimage}
\end{center}
\end{figure}
Hard-sphere colloidal systems exhibit a glass
transition at which the dynamics becomes very slow  around a
volume fraction of  ${\phi}_{glass} = 0.58$ \cite{pusey}. 
 We use
poly-methylmethacrylate (PMMA) particles, of radius,
$\sigma=1.3\mu$m, which are dyed with rhodamine and are sterically
stabilised to prevent aggregation. The rhodamine dye makes it
possible to follow the Brownian motion of all the 
individual particles on a plane using a confocal microscope.
There is a polydispersity of about $5\%$ in the
particle size to prevent crystallization in the system.  We use a
mixture of cyclohexylbromide and decalin as our solvent, to match
the density and index of refraction of the particles. The
constituents and preparation procedure leads to particles that are
indistinguishable from hard spheres. To verify this, we have
measured the crystallization density, which is a very sensitive
measure for deviations from hard-sphere behavior, and found it to
agree to within a fraction of a percent to that of true hard
spheres.

Two-dimensional images were acquired in a field of view of 100
$\mu$m x 100 $\mu$m.  The time interval between each image frame
is $0.05$ sec, which is approximately $1/10^{th}$ of the Brownian
timescale $\tau_{B}, \tau_{B} = \eta d^{3}/k_{B}T \approx 0.75
sec$, where $\eta $ is the solvent viscosity of the suspension.
Following around 2000 particles of radius $1.3\mu m$  in real
time using a fast confocal microscope (Zeiss LSM live) allows us
to reconstruct all particle trajectories (Fig. \ref{Fig_2Dimage}).
\begin{figure}
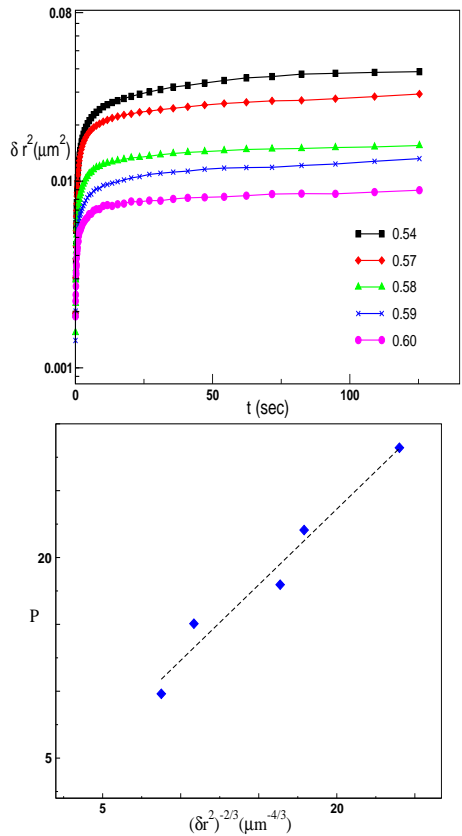

\includegraphics[width=6cm,height=5.5cm]{SDMSD.eps}
\includegraphics[width=5.5cm,height=5.5cm]{Nfit_pres.eps}
\caption{LEFT: Mean square displacement versus time at different volume fractions
$\phi = 0.54, 0.57, 0.58, 0.59, 0.60$. RIGHT: The pressure, as obtained
in terms of the volume fraction from the empirical
hard-sphere relation, plotted as a function of the plateau value
of the mean-square displacement. The full line is a $P^{3/2}$ fit.}
\label{MSD0}
\end{figure}

The mean square displacement per particle $<(\delta r)^{2}>$ as a
function of time  shows a  plateau, indicating that in our
experimental time-window, each particle moves in a `cage'
constituted by its neighboring particles.
The plateau value  decreases with increasing volume
fraction of the suspension, because the particle motion is more
restricted (Fig. \ref{MSD0}).
 Clearly, measurement
times have to be long enough for soft modes (if present) to be visible,
 yet short enough that the system  stays within a
basin.  The very existence of a plateau in
our measured mean-squared displacement shows that there is a range
of time scales for which local equilibrium can be justified and
the soft modes can be measured, without many activated events taking place.

If there is a proliferation of soft modes above a   gap, the value of the plateau in the
mean-squared displacement, as Brito and Wyart \cite{Brito2}
showed, is dominated by the modes just above the gap. An estimate of the gap value then leads~\cite{Brito1, Brito2}
to the prediction  that the plateau value of the mean-squared
displacement scales with pressure as $p^{-3/2}$. This explicit
scaling can be tested directly in our experiments by plotting
the value of the plateau in the mean square displacement curve vs.
the pressure; the latter can be computed from the volume fraction
using an empirical equation of state for the colloidal hard
spheres, for example \cite{Liu}, $ p = - (k_BT/v)\phi^2
\frac{d}{d\phi}ln[ ((\phi_{max}/\phi)^{(1/3)}-1)^3 ]$, with $v$
the volume per particle. Fitting the data, it is clear that they
are indeed compatible with an exponent~${-3/2}$ (Fig. \ref{MSD0}).
If we admit that the system is in local equilibrium during the
time-interval associated with the plateau, then the values of the
squared-displacements are independent of the dynamics, which may
go from  overdamped to purely ballistic. (What {\em does} depend on the 
nature of the dynamics is the actual time-dependence of the displacements,
but not their statistical distribution).

To obtain the Density of States (DOS), we can compute the normal
modes of the (not necessarily harmonic) vibrations in the cages
from the displacement correlations. Denoting $u_a(t)$ the components of the
particle displacements from the average position along
the confocal plane $u_{a}(t)= \{(x_i(t)-\langle x_i \rangle),(y_i(t)-\langle y_i \rangle)\}$,
we obtain the displacement correlation \cite{Jorge, Hess, Harris}
matrix (of dimension twice the number of observed
particles) as,
\begin{equation}
D_{ab} = \langle u_a(t) u_b(t) \rangle,
\end{equation}
where $\langle \bullet \rangle$ denotes average over the time of
the whole period of measurement, which is about $175 sec$.
Diagonalizing $D_{ab}$ we obtain the normal modes, and the
corresponding eigenvalues $\lambda_{a}$.
We express the results in terms of the relevant quantities,
\begin{equation}
 \omega_{a} = \sqrt{1/\lambda_{a}} ,
\end{equation}
which are the temporal frequencies the
system would have {\em if it were harmonic}. 
Let us stress that we do not assign any meaning of temporal vibrational  frequency
to $\omega_a$ at this stage.

\begin{figure}
\begin{center}
\includegraphics[width=5cm,height=5cm]{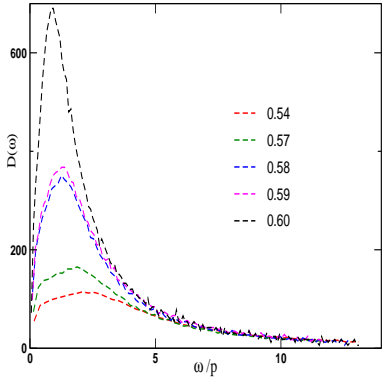}
\caption{Rescaled Density of states at different volume fractions,
$\phi = 0.54,0.57,0.58,0.59,0.60$. The frequencies along
the horizontal axis are  scaled by the pressure (estimated from the volume fraction)
times $(2\sigma)^{1/3}$. The Y-axis is  scaled
to overlap the high frequency tails of the distributions.
The spectra are consistent with the hypothesis that the curves have a low-frequency gap
that closes as the pressure grows, and approach a universal envelope curve as the pressure
goes to infinity.}
\label{spDOS}
\end{center}
\end{figure}

Let us now turn to the results DOS. Because ``hard" modes are
expected to have eigenvalues proportional to the  pressure, we
scale out this trivial effect by plotting the DOS in terms of
$\omega/p$. The rescaled density of states shows a pronounced peak
with increased (Fig. \ref{spDOS}) volume fraction in agreement with
the hard sphere theory~\cite{Brito2,Jorge}.

\begin{figure}
\begin{center}
\includegraphics[width=6cm,height=6cm]{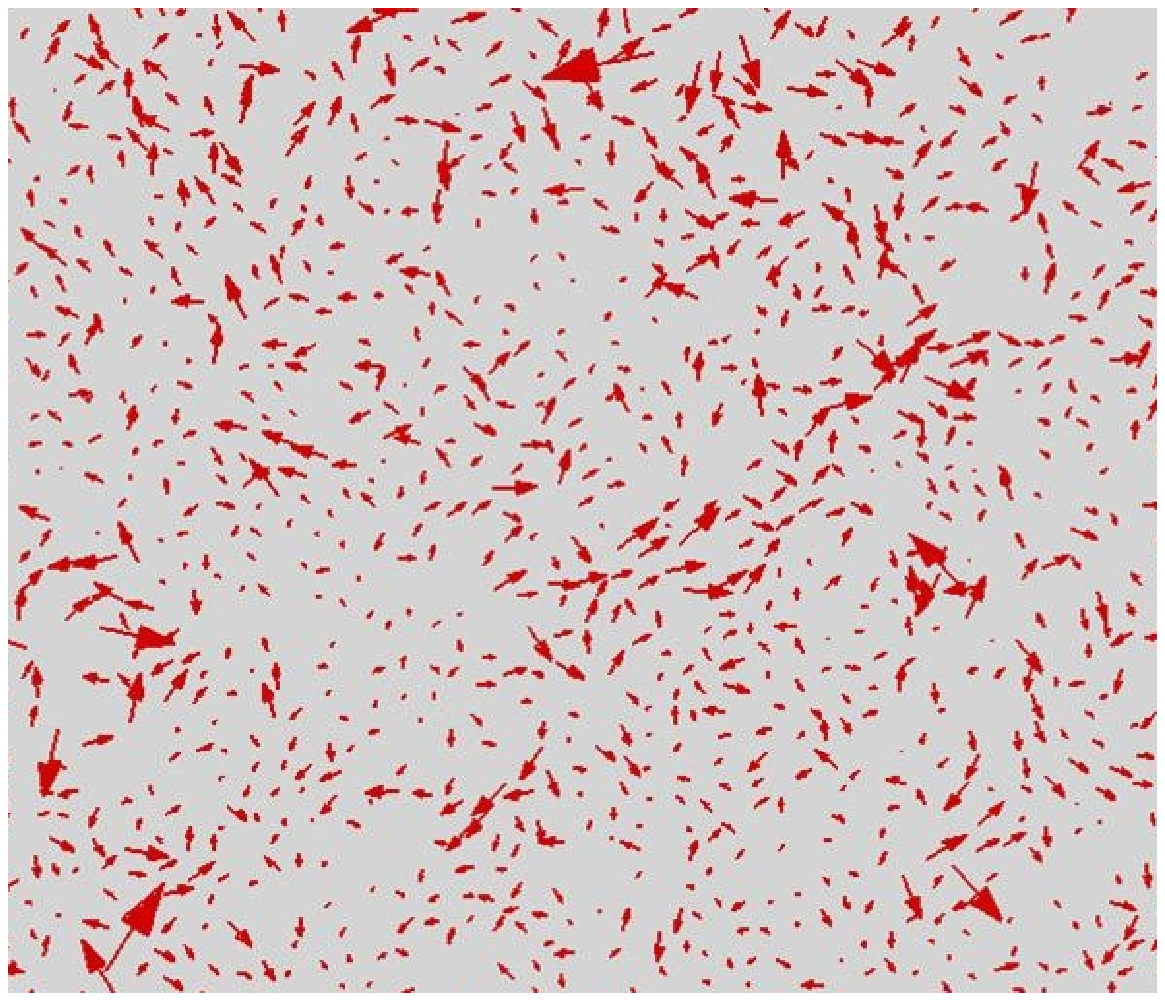}
\quad
\quad
\vspace{3mm}
\includegraphics[width=6cm,height=6cm]{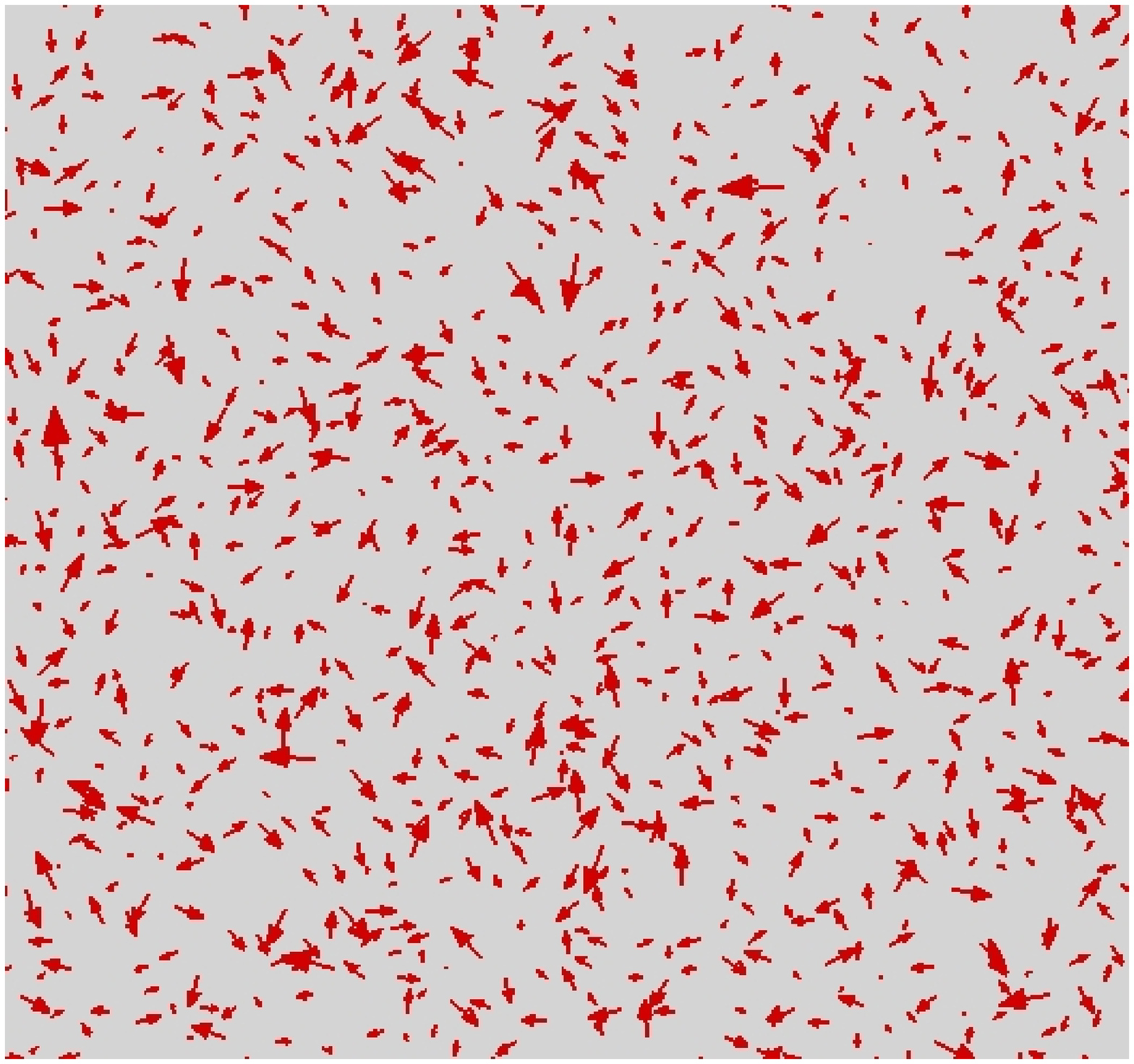}
\caption{Two eigenmodes for volume fraction 0.60. LEFT: low $\omega$: collective ``swirling" motion is visible.
RIGHT: high $\omega$ motion has short range correlations. }
\label{eigen}
\end{center}
\end{figure}

\begin{figure}[!htbp]
\begin{minipage}{7cm}
\includegraphics[width=6cm,height=6cm]{Avcorr-9.eps}
\end{minipage}
\hspace{2mm}
\begin{minipage}{7cm}
\includegraphics[width=6cm,height=6cm]{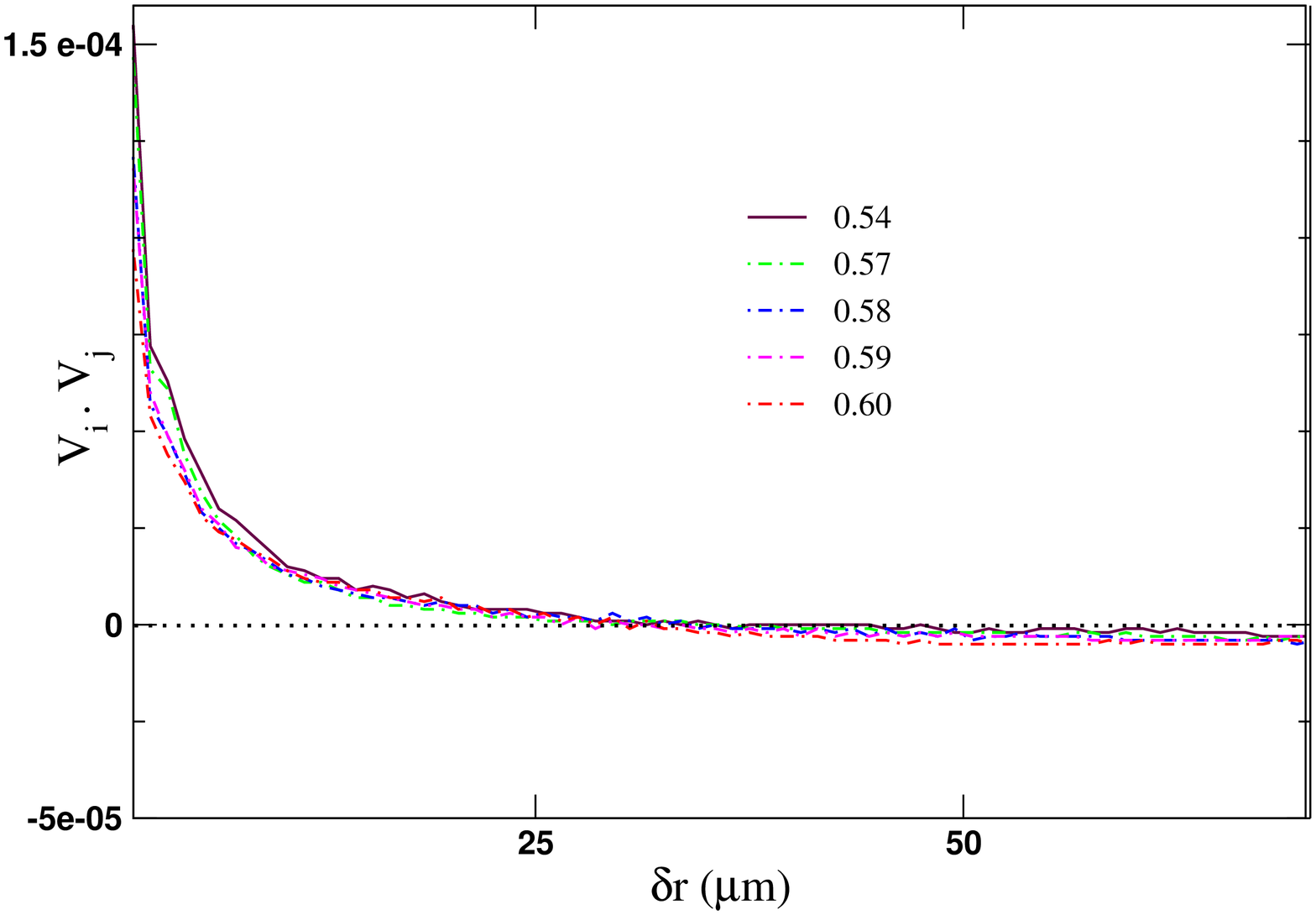}
\end{minipage}
\caption{An analysis of eigenmodes:  the  correlation
function $\vec{v_i} \bullet \vec{v_j} $ of displacements
$\vec{v_i}$ in a single mode (cfr Figure \ref{eigen}), versus the
distance $\delta r = |\vec{r_i}-\vec{r_j}|$ between particles. The
correlations are computed for eigenmodes in different parts of the
spectrum, as shown in $D(\omega)$ in the inset by the respective
colored lines. Soft modes have  a long correlation tail,
characterizing the ``swirling" motion. The hardest modes are made
of pairs of particles moving in anticorrelated manner. RIGHT:
``soft"  modes corresponding to systems at different packing
fractions. The general shape of the modes seems very weakly
dependent on the packing fraction. The volume fraction is $0.60$.}
\label{corr}
\end{figure}

What do the soft modes look like ?  The two dimensional eigenvector
corresponding to single soft modes shows the emergence of rather
large vortex-like structures (Fig. \ref{eigen}). On  the
contrary, for higher eigenvalues the eigenvector field looks
random.  To quantify the order in the low-frequency
eigenvector field, we calculated the orientational correlation
function defined as the scalar product $ (\vec {v}_{i} .
\vec{v}_{j})$,  where $\vec{v}_{i}$ and $\vec{v}_{j}$ are the
two-dimensional eigenvector components corresponding to
the $i^{th}$ and $j^{th}$ particle for a single eigenmode.
The result for eigenmodes in the different parts of the spectrum
are shown in Fig. \ref{corr}.
Softer  modes involve  motion correlated over many
(tens of) interparticle distances, while the hardest modes are
essentially pairs of particles vibrating with opposite phases. The
correlations characterizing soft modes are very weakly dependent
on pressure (Fig. \ref{corr}), suggesting that this correlation
length stays finite in the large-pressure limit.

{\bf Mode-tomography:}
In our experimental setup we measure the displacements of
particles in a small and two-dimensional slice of the real system,
and we diagonalise a submatrix of the full correlation matrix.
This is not only due to an experimental, but also to a numerical
limitation: diagonalizing the full correlation matrix is an
impossible task. The problem is not hopeless, because our slice is
typical of all others and will give statistically equal results.
In particular, the mean squared displacement  per particle and per
spatial dimension is the same when restricted to a slice and in
the whole system. One can say more: consider an eigenvector
$\tilde V $ with eigenvalue $\tilde \lambda$ of the restricted
covariance matrix,  involving only the measured particles and
their motion along the confocal plane.  The eigenvectors of the
full system $\{V_{a}\}$ form a complete basis so we can expand
$\tilde V$ in terms of $V_a$ as:
\begin{equation}
\tilde V = \sum_a \; c_a \; V_a\;\;\; \;\;\;\;\;\; \;\;\;with
\;\;\; \;\;\;\;\;\; \;\;\;\sum_a \;
c_a^2=1\label{pp}.\end{equation} One has $\tilde \lambda = \sum_a
c^2_a \lambda_a$, which can also be written as $\sum_a \; c_a^2 \;
(\tilde\lambda -\lambda_a)=0$. It is clear from this and the fact
that eigenvalues are positive  that the existence of a large
eigenvalue $\tilde \lambda$ for the restricted problem implies the
existence of a large eigenvalue $\lambda_a \ge \tilde \lambda $
for the complete problem, or, equivalently, $\omega_a \le \tilde
\omega$. Each soft mode of the restricted problem implies the
existence of --  and has a
    large projection
  over -- soft modes of the complete system.
In particular, a
mode with $\tilde \omega=0$ in the restricted problem implies a
mode $\omega=0$ in the true system.
Although we cannot prove at this point  that the spectrum
 associated with a large two-dimensional slice
coincides exactly with that of the full system, the soft modes we
observe are indeed a reflection of all the soft modes of the full
system.

Another argument that justifies the observation on a restricted slice is that
the vibration  we observe in our slice is induced by the global soft modes. As mentioned above, 
the  DOS
 $D(\omega)$  has a sharp peak at a gap value  $\omega^*(p)$,
below which there are very  few modes ($\omega^*(p)$  goes to zero as $p \rightarrow \infty$).
 The   mean squared displacement is dominated  at high pressures  by the peak value
contribution~\cite{Brito2}   $<(\delta r)^{2}>  \sim
\frac{D(\omega^*)}{\omega^*}$.  {This means that at large pressure, 
{\em essentially  all} the motion of each particle  is given by  a combination of
the soft modes just above the gap,  i.e. those having  $\omega \sim \omega^*(P)$.}
\\

In conclusion, we have observed an excess of low frequency modes
in colloidal hard sphere system around the glass transition
volume fraction. These modes   show large-scale correlations in the velocities of particles,
 extending over many particle diameters.
The existence of  soft modes had been predicted for ideal  hard spheres
on the basis of theoretical considerations.
Our  experimental study shows  that they do exist colloidal suspensions,  
 probably the most studied system
in the field of glassy dynamics. These vibrational modes signal the onset
of macroscopic elasticity and give a microscopic insight in the collectivity of the particle dynamics near the glass transition.

We have argued that the normal modes of a restricted set of particles gives information on the whole system.
Thanks to this, a well established technique such as  confocal microscopy can be used to
study global vibrational properties of the system, which contain
very detailed information on the geometry
of the configurations.
The above strategy can be applied for much more complex colloidal
systems. Following  particles for longer times will further allow us to assess the relevance of the
low frequency soft modes to long time, activated dynamics near glass transition.


\end{document}